\documentclass[letter]{aa} 

\usepackage{graphicx}
\usepackage{txfonts}
\usepackage{hyperref}

\begin{document}

\title{Messier 90 galactic halo rotation in the microwaves}


\author{F. De Paolis \inst{1,2,3} \fnmsep\thanks{Email: depaolis@le.infn.it}
        \and
        V.G. Gurzadyan\inst{4,5}
        \and
        A.L. Kashin\inst{4}
        \and
        G. Yegoryan\inst{4}
        \and
        A. Qadir\inst{6}
        \and
        N. Tahir\inst{7}
        \and
        Ph. Jetzer
 }

\institute{Department of Mathematics and Physics {\it ``E. De Giorgi''} , University of Salento, Via per Arnesano, CP-I93, I-73100, Lecce, Italy
           \and
           INFN, Sezione di Lecce, Via per Arnesano, CP-193, I-73100, Lecce, Italy
           \and
           INAF, Sezione di Lecce, Via per Arnesano, CP-193, I-73100, Lecce, Italy
           \and
           Center for Cosmology and Astrophysics, Alikhanian National Laboratory and Yerevan State University, Yerevan, Armenia 
              \and SIA, Sapienza University of Rome, Rome, Italy 
              \and
              Pakistan Academy of Sciences, Constitution Avenue, G-5, Islamabad, Pakistan
              \and
              Department of Physics and Astronomy, School of Natural Sciences (SNS), National University of Sciences and Technology (NUST), Sector H-12, 44000, Islamabad, Pakistan
              \and 
              Physik Institut, Universität Zürich, Winterthurerstrasse 190, CH-8057 Zürich, Switzerland
}

\date{Received XXX; accepted ZZZ}

\abstract{
We used {\it Planck} data to study the Virgo Cluster's galaxy M 90 and its surroundings. We find, as in the case of certain galaxies of the Local Group and its vicinity,  a substantial temperature asymmetry that probably arises from the rotation of the M 90 halo and extends up to about one degree from its centre. This temperature asymmetry is particularly remarkable as M 90 is a rare blueshifted galaxy of the Virgo Cluster, and it thus has possible implications for the cluster internal dynamics versus the galactic halo's formation and structure. Possible explanations for the observed effect are discussed.}


\maketitle

\section{Introduction}
The baryonic content of galactic halos, in the form of hot gas in the circumgalactic medium but also with contributions from cooler gas and dust, plays a fundamental role in our understanding of the dynamics and evolution of galaxies. While galactic halos are often less visible than the galaxies they surround, they contain a significant fraction of the baryonic matter in the Universe and are key to understanding the complex interplay between galaxy formation, feedback, and the cosmic environment.

In a series of papers published in the last decade, WMAP (Wilkinson Microwave  Anisotropy Probe) and {\it Planck} data were used with the aim of tracing  the galactic halos of some nearby edge-on spiral galaxies belonging to the Local Group and its vicinity, such as M 31 (\citealt{depaolis2011}, \citealt{depaolis2014}), Centaurus A \citep{depaolis2015}, M 33 \citep{depaolis2016}, M 104 \citep{depaolis2019},
M 82 \citep{gurzadyan2015}, M 81 \citep{gurzadyan2018}, and other galaxies (see also \citealt{tahir2022} and references therein). In all those cases, a clear asymmetry in the microwave temperature was detected that extended up to several times the optical size of the galaxies, probably induced by the Doppler effect due to the galactic rotation. Additionally, {\it Planck} data sometimes indicated relatively complex galactic halo dynamics, as also suggested by other studies.

In this Letter we extend the analysis previously limited to nearby edge-on galaxies to the case of a relatively far galaxy: Messier 90 (M 90), the brightest  late-type galaxy in the Virgo Cluster.
It is a particularly interesting and,  for several reasons, almost unique galaxy. Spectral observations of M 90 show it is moving towards the Milky Way despite the overall redshift of the Virgo Cluster. The spectrum of M 90 is actually blueshifted, indicating that it is moving towards us with a heliocentric radial velocity of  $\simeq -235\pm 4$ km\,s$^{-1}$ (or a galactocentric radial velocity of $\simeq -282\pm 4$ km\,s$^{-1}$), corresponding to a redshift $z=-0.000784\pm 0.000013$ (see e.g. \citealt{boselli2016}). The blueshift was originally used to argue that M 90 was actually an object in the foreground of the Virgo Cluster. However, since the observation was limited mostly to galaxies in the same part of the sky as the Virgo Cluster, it is further evidence of the large range in velocities of objects within the Virgo Cluster itself.

The spiral arms of M 90 rotate at a significant speed, and the galaxy exhibits a typical rotation curve: the velocity (based on HI kinematics data) peaks at a value of about 250 km s$^{-1}$ and remains relatively constant at increasing distances from the galactic centre.
M 90 is also notable for its relatively low star formation rate. It is thought to  be an anemic `galaxy', a term coined by van der Bergh  \citep{vabdenbergh1976,vabdenbergh1991} to classify galaxies in an intermediate form between gas-rich star-forming spiral galaxies and gas-poor inactive lenticulars. These objects are characterised by a rather low contrast between their spiral arms and the disc. Galaxies in rich clusters are thought to evolve into anemic galaxies, and the Milky Way is expected to follow this trajectory (another example is NGC 4921 in the Coma cluster).
\begin{figure}
\centering
\hspace{-0.2cm}
  \includegraphics[width=0.5\textwidth]{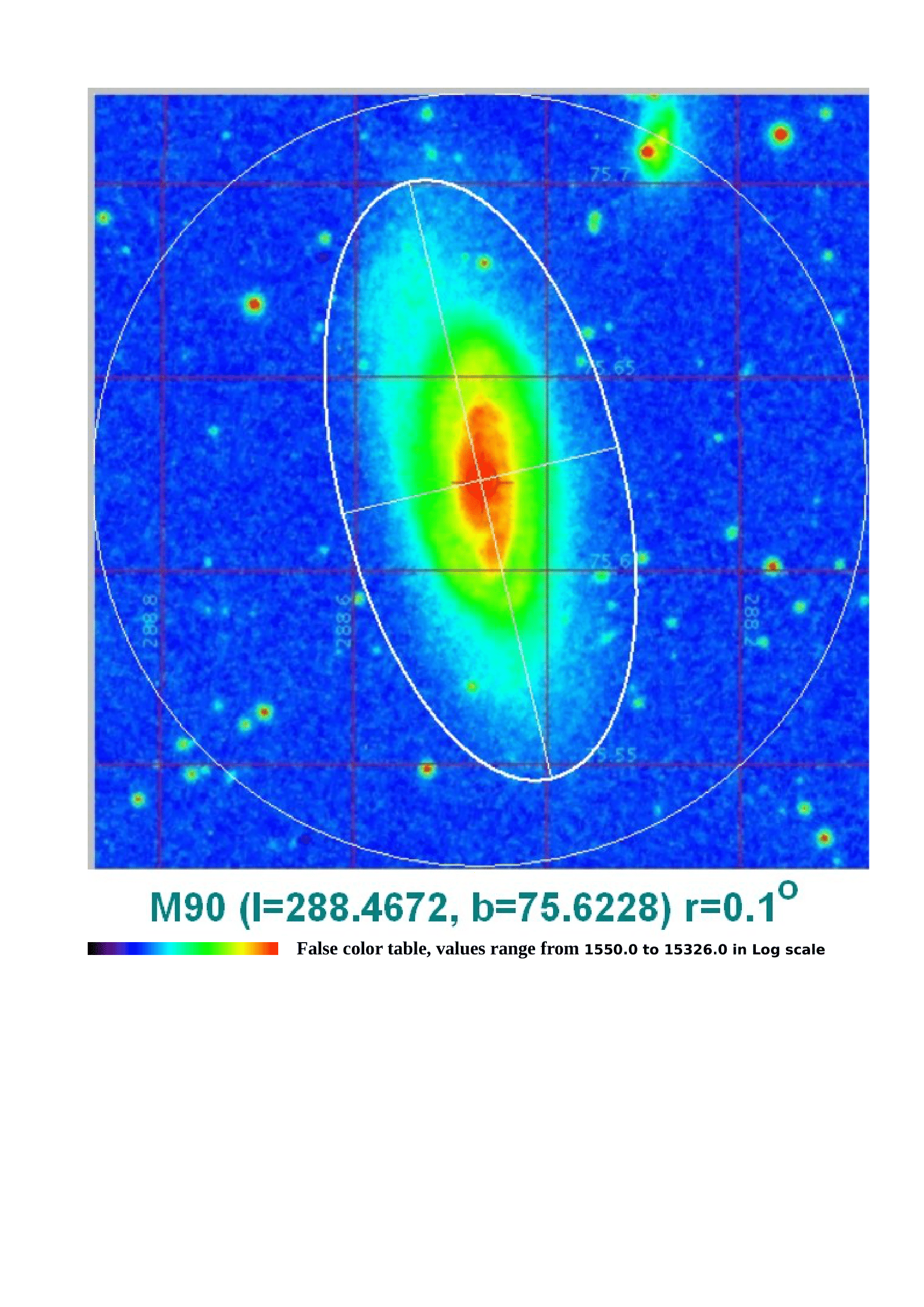}
  \caption{Optical Sloan Digital Sky Survey image centred on M 90 in false colour (taken from the web site https://skyview.gsfc.nasa.gov/current/cgi/query.pl). The optical extension of M 90 is indicated by the ellipse with major and minor axes of  $9.5\arcmin$ and $4.4\arcmin$, respectively. The circle has a radius of  $0.1 \degr$.} \label{fig1}
\end{figure}
M 90 is also characterised by unusual rotation dynamics, which is thought to result from its interactions with the Virgo Cluster environment. M 90 may be anemic since, in the course of its orbit, about $3\times 10^8$ years ago it passed close to the centre of the Virgo Cluster. The density in the gas in the inner Virgo Cluster
weighed on M 90 like  a strong headwind, stripping an enormous quantity of gas from the galaxy and creating a diffuse halo that can be seen around it. 
 \cite{boselli2016} used MegaCam at the CFHT (Canada-France-Hawaii Telescope) to obtain a deep narrow-band H$\alpha +$ [NII] wide-field image of M 90, which revealed the presence of  long tails of diffuse, ionised  gas extending up to a projected distance of about 145 kpc from the galactic disc (corresponding to an angular extension of $\simeq 0.5\degr$). Such features clearly indicate that M 90 is undergoing a 
ram-pressure stripping event.  This stripping event should have enriched the region beyond $0.5\degr$ with gas and dust from the galactic centre. The ram-pressure stripping may also explain the gas removal from the internal regions of M 90 and, consequently, the quenching of the star formation activity that can be observed  in many galaxies located in high-density regions, often surrounding the centre of galaxy clusters.

These are some of the reasons why we selected M 90 to be analysed using {\it Planck} data with the aim to check if its halo shows some temperature asymmetry in the microwaves. 

\section{The M 90 galaxy}
\subsection{M 90: Generalities}
The galaxy M 90 (also designed as NGC 4569, UGC 7786, and Arp 76), with centre at coordinates RA: 12$^{\rm h}$ 36$^{\rm m}$ 49.8$^{\rm s}$, Dec: $13\degr$ $09\arcmin$ $46\arcsec$, corresponding to galactic coordinates $l=288.4672\degr$ and $b=75.6228\degr$ (see Fig. \ref{fig1} for an optical Sloan Digital Sky Survey  view), is a spiral galaxy located about $1.7\degr$ from the M 87 galaxy, which itself lies at the centre of the Virgo galaxy cluster (see Fig. 1 in \citealt{boselli2016}). 
M 90 is classified as an SA(rs)bc galaxy, meaning it is a spiral galaxy with a somewhat loosely bound structure and no bar (SA), an inner ring (rs), and moderately developed spiral arms (bc). The galaxy has a bright central bulge, but its spiral arms are less well defined in comparison to other spirals. The distance to M 90 is estimated to be $18.0^{+0.9}_{-0.6}$ Mpc, assuming a Hubble constant of  $68$ km s$^{-1}$\,Mpc$^{-1}$  (note that \citealt{tschoke2001} obtained a distance of about 16.8 Mpc using a Hubble constant value of  $75$ km s$^{-1}$\,Mpc$^{-1}$). The real distance to M 90 is ambiguous and determining it will require dedicated studies with reliably calibrated methods.

\subsection{Planck data analysis towards M 90}
\begin{figure}
 \centering
  \hspace{-0.2cm}
  \includegraphics[width=0.5\textwidth]{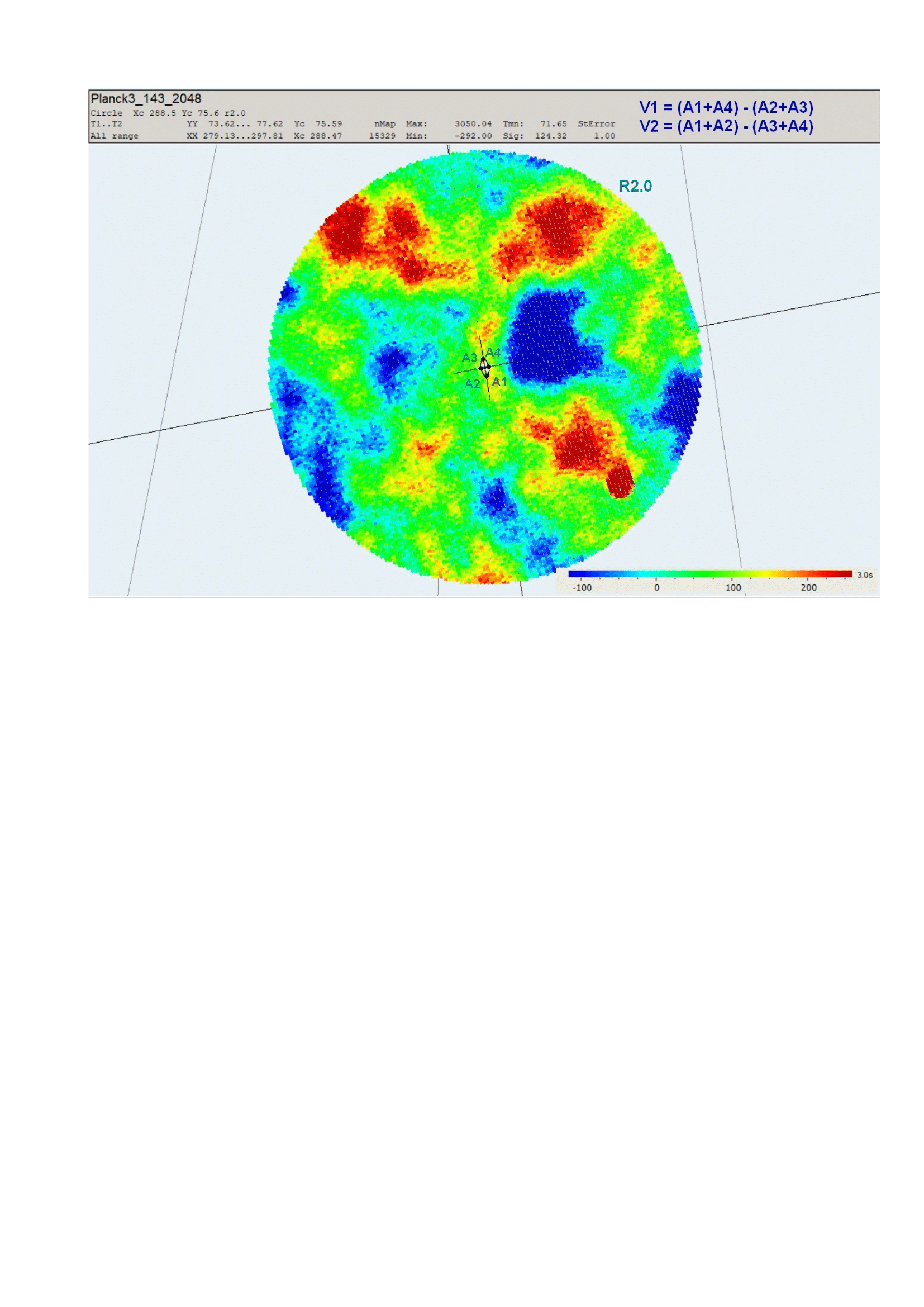}
   \caption{{\it Planck} view in the 143 GHz band of the region around  M 90 (with $2.0\degr$). The quadrants labelled A1, A2, A3, and A4 (going clockwise starting from A1 in the bottom right) were used for the temperature analysis (see the main text for details).} \label{fig2}
\end{figure}

To analyse {\it Planck} data  with the aim of searching for a temperature asymmetry towards the halo of M 90, we adopted the same procedure described in the previous papers. Considering the publicly released {\it Planck} 2018 data, \footnote{  {\it Planck} Legacy Archive: http://pla.esac.esa.int} as described in \cite{agha2020}, we used the bands at 70 GHz, 100 GHz, and 143 GHz, as well as the foreground-corrected SMICA map (indicated as SMICA\_H).
In Fig. \ref{fig2} the region around M 90 is shown in the 143 GHz {\it Planck} band. The region extends up to $2\degr$ around M 90. 
An angular distance of $1\degr$ at the M 90 distance corresponds to about 300 kpc.

The quadrants labelled A1, A2, A3, and A4 were used for the temperature analysis in two different configurations, variant 1 (V1) and variant 2 (V2).
Similar to other nearby galaxies analysed in some previous papers, M 90 shows a temperature asymmetry in
{\it Planck} data. This effect is clearly visible in all considered {\it Planck} bands (70, 100, and 143 GHz) as well as
in the foreground-corrected SMICA map. 
In V1 (see the upper panel of Fig. \ref{fig4}) the temperature asymmetry is clearly visible in the region within $0.5\degr$, where it amounts to about 42-43 $\mu$K in all bands, while the control fields (lower panel in Fig. \ref{fig4}) do not show  any deviation from expectations. In the region within $1\degr$, the temperature asymmetry diminishes slightly but remains substantial, amounting to about 35 $\mu$K. Only in the outer region does the temperature asymmetry disappear.

For V2 (see Fig. \ref{fig5}), the temperature asymmetry is practically null in the $0.5\degr$ region, while it increases substantially in the region within $1\degr$, amounting to about 55 $\mu$K in all bands. This clearly means that the effect is induced by the region between about 150 and 300 kpc around M 90. The temperature asymmetry then disappears in the $1.5\degr$ region.

The fact that the temperature asymmetry always appears to be frequency-independent indicates that it may be due to gas and/or dust emission in the M 90 halo and is modulated by its rotation (other possible explanations are discussed in the next section). The effect clearly extends up to about $1 \degr$ from the M 90 centre, corresponding to a distance of about 300 kpc. 
It goes without saying that the detected temperature asymmetry is much more extended than the visible M 90 disc. 
The temperature asymmetry practically disappears in the  $1.5 \degr$ region, as can be clearly seen in Figs. \ref{fig4} and \ref{fig5}. 
\begin{figure}
 \centering
  \hspace{-0.4cm}
  \includegraphics[width=0.51\textwidth]{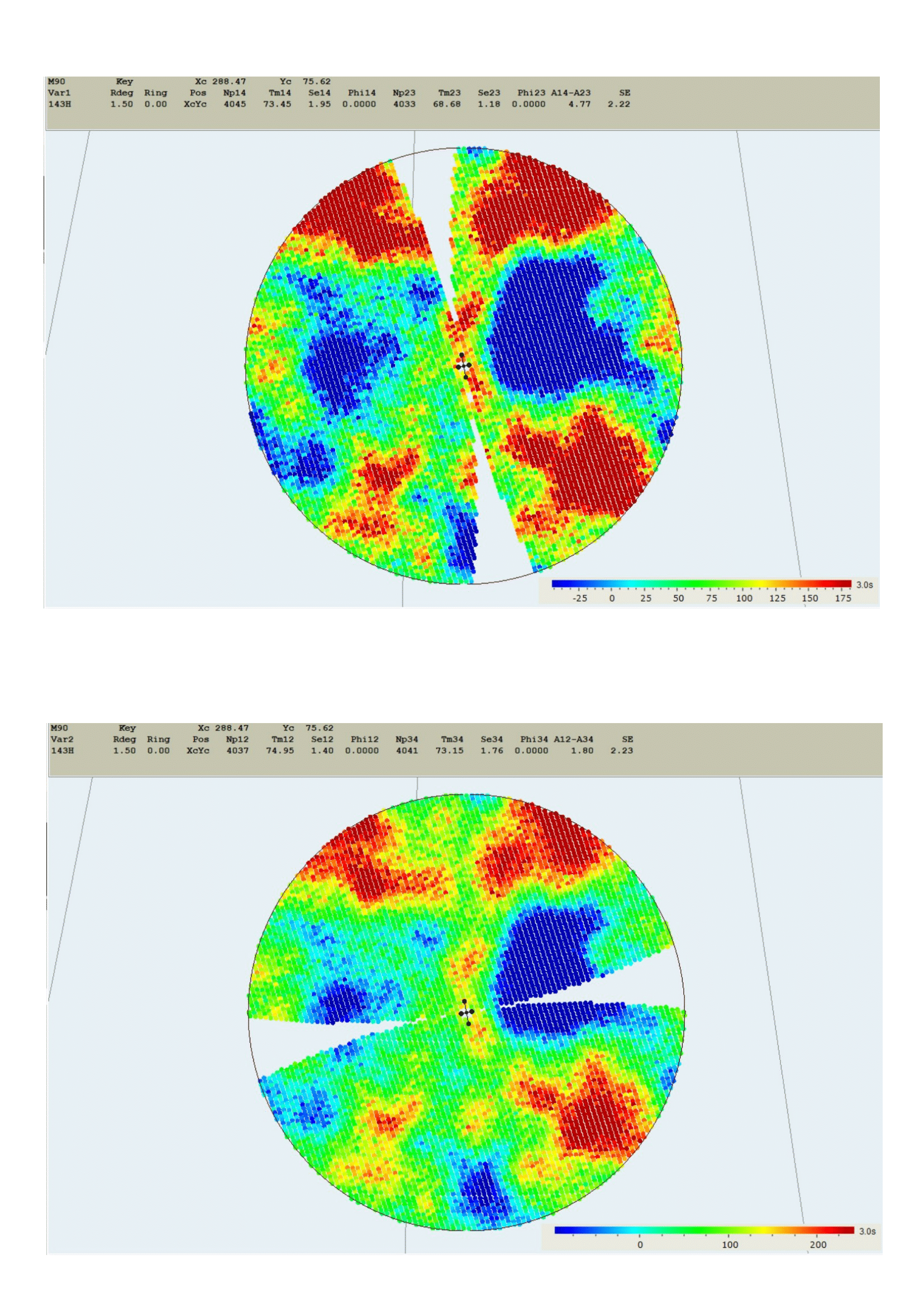}
   \caption{Upper panel: {\it Planck} map at 143 GHZ of quadrants $A1$ and $A4$ (right) and $A2$ and $A3$ (left), as indicated in Fig. \ref{fig2}. This panel corresponds to V1. Bottom panel: Same as the upper panel but for V2. Regions $A1$ and $A2$ are shown in the bottom part of the panel and regions $A3$ and $A4$ are in the upper part.} 
   \label{fig3}
\end{figure}

For a model in which the temperature asymmetry derives from the emission induced by the rotation of the M 90 galactic halo, one expects that
\begin{equation}
\frac{\Delta T}{T}=2\frac{V_{\rm rot}}{c}\tau_{\rm eff}\sin i
\label{eq1}
,\end{equation}
where $\Delta T$ is the measured temperature asymmetry between the two considered galaxy lobes and the $T$ in the denominator is the cosmic microwave background temperature; $\tau_{\rm eff}$ is the effective gas/dust  cloud optical depth, which depends on the cloud filling factor and the averaged optical depth within  a {\it Planck} band (see e.g. \citealt{depaolis2014}); and $i$ is the inclination angle between the galaxy rotation direction vector and the line of sight to the galaxy.
Taking into account that the rotation velocity, $V_{\rm rot}(R),$ within a galactocentric distance $R$ is related to the galaxy dynamical mass, $M_{\rm dyn}(R),$ as 
\begin{equation}
    V_{\rm rot}(R)=\sqrt{\frac{G M_{\rm dyn}(R)}{R}}
    \label{eq2}
,\end{equation}
one can derive the following relation for the  M 90 dynamical mass:
\begin{equation}
  M_{\rm dyn}(R)\simeq 7\times 10^4 \left(\frac{R}{100\,\, {\rm kpc}}\right)\frac{\Delta T^2_{\mu{\rm K}}}{(\tau_{\rm eff}\,\sin i)^2} \, M_{\odot}
    \label{eq3}
.\end{equation}
Since typical values estimated through numerical simulations for $\tau_{\rm eff}$ are about $10^{-3}-10^{-2}$ (see \citealt{tahir2019} for an estimation of the $\tau_{\rm eff}$ for the M 31 galaxy, which looks not so different from M 90), one can derive an estimate for the $M_{\rm dyn}$ of M 90:
\begin{equation}
  M_{\rm dyn}(300\, {\rm kpc})\simeq 10^{13} \, \,M_{\odot}
    \label{eq4}
.\end{equation}
This is in substantial agreement with literature data for the total mass amount  of M 90  within about 300 kpc (see e.g.  \citealt{haan2008,boselli2016} and references therein).
\begin{figure}
\centering
  \hspace{-0.5cm}
  \includegraphics[width=0.49\textwidth]{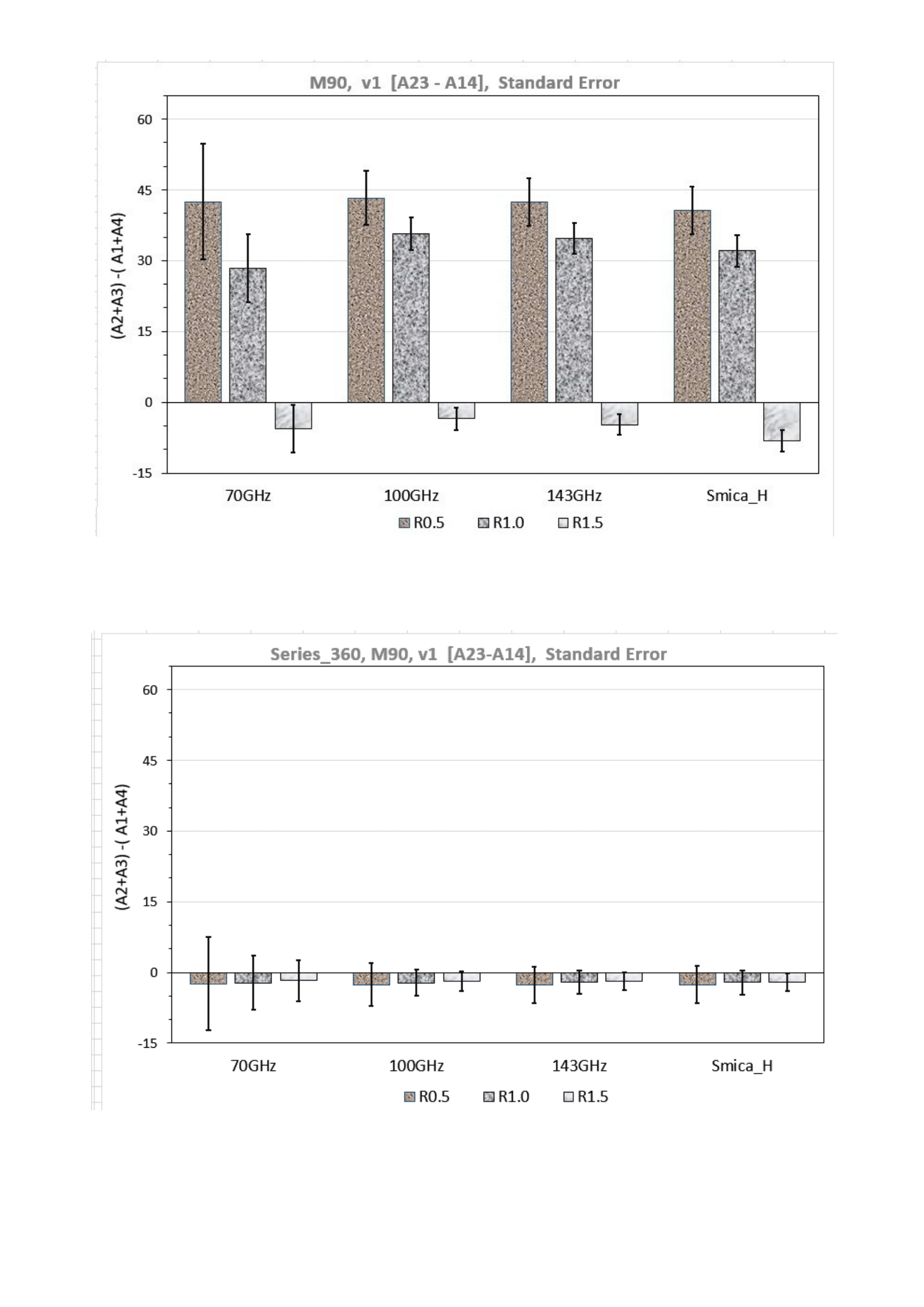}
   \caption{Upper panel: Temperature asymmetry towards M 90  (in $\mu$K, with the standard errors) of V1  (A2+A3)-(A1+A4) in the four considered {\it Planck} bands  at 70 GHz, 100 GHz, 143 GHz, and in the SMICA$_H$  band. The temperature is measured  within $0.5\degr$, 1$\degr$, and $1.5\degr$. Bottom panel: Same as the upper panel but for 360 control fields  spaced by one degree from each other in Galactic latitude and at the same latitude as M 90.} \label{fig4}
\end{figure}
\begin{figure}
 \centering
  \hspace{-0.8cm}
  \includegraphics[width=0.49\textwidth]{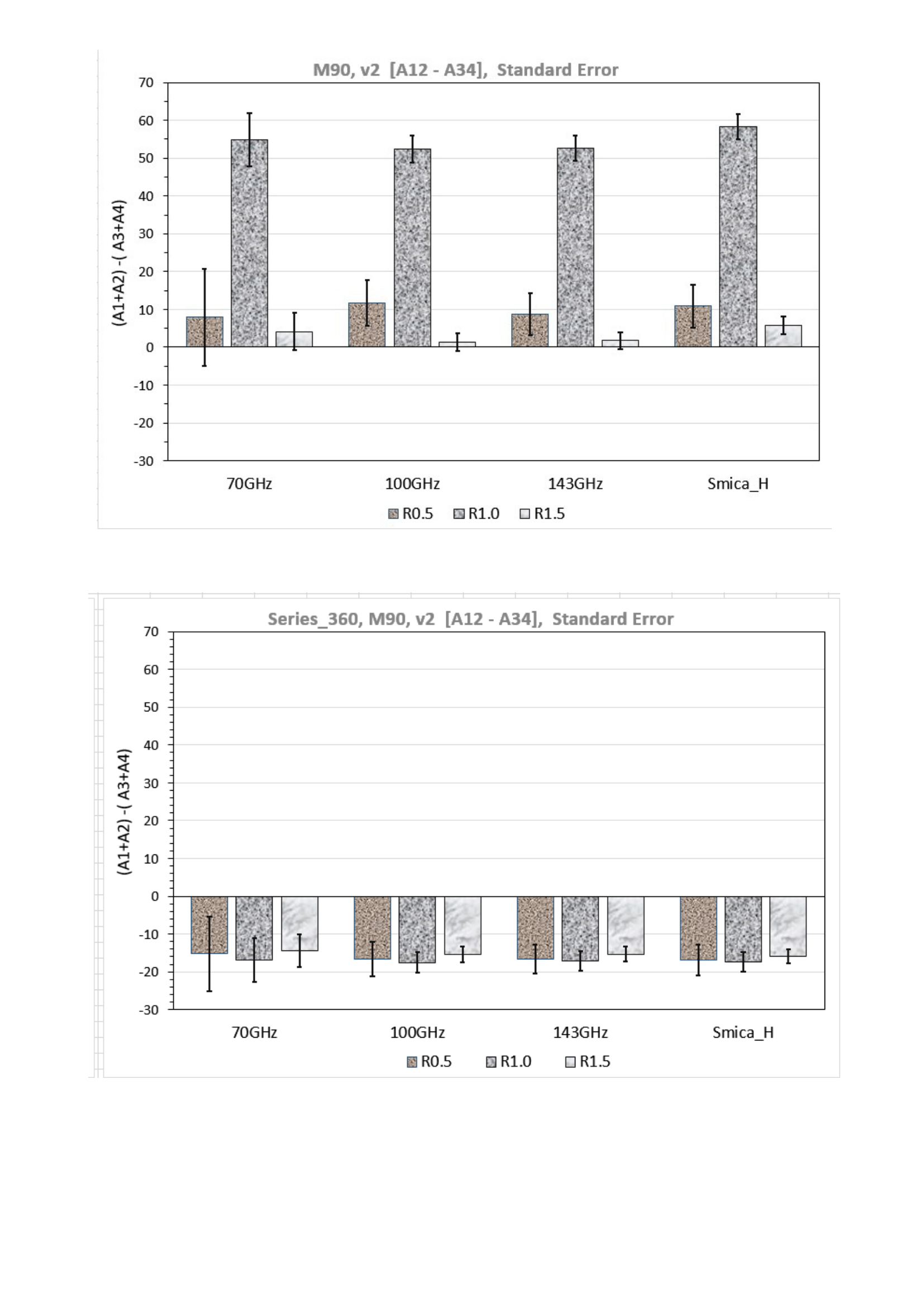}
   \caption{Upper panel: Temperature asymmetry towards M 90 (in $\mu$K; with the standard errors) of V2  (A1+A2)-(A3+A4) in the four considered {\it Planck} bands. The temperature is measured  within $0.5\degr$, 1$\degr$, and $1.5\degr$. Bottom panel: Same as the upper panel but for 360 control fields   spaced by one degree from each other in Galactic latitude and at the same latitude as M 90.} \label{fig5}
\end{figure}
We note that there is no useful information in the literature regarding the rotation of the M 90 galactic halo, unlike for other galaxies of the Local Group analysed previously (such as M31, Cen A, and M 33) for which spectroscopic data of either globular clusters or HI gas are available in the literature. Such data would be extremely useful for placing priors on the rotation amount and direction of the M 90 galactic halo.

\section{Conclusions}
Both adopted variants show a clear and substantial temperature asymmetry in the {\it Planck} data towards M 90, extending up to about 300 kpc and amounting to about 50 $\mu$K. It is important to note that the fact that M 90 lies at a very high Galactic latitude implies that a contamination by clouds and  cirruses of the Milky Way (as hypothesised by \citealt{ade2015} to explain the temperature asymmetry observed towards the Andromeda galaxy)  is extremely unlikely.

Regarding the origin of this temperature asymmetry, in general, five  different emission mechanisms might account for it: (i) free-free emission produced by a hot plasma surrounding M 90 (see e.g. \citealt{sun2010}),
(ii) synchrotron emission produced by magnetic fields in the M 90 halo on fast moving electrons (see e.g. \citealt{dolag2000}), (iii) anomalous microwave emission from dust grains (see e.g. \citealt{ade2011,leitch2013,dickinson2018}), (iv) the rotating kinetic Sunyaev-Zeldovich (rkSZ) effect from ordered gas rotation (see e.g. \citealt{chluba2001,chluba2002,baldi2017,matilla2020}), and (v) cold gas
clouds populating the halo of M 90 (as first proposed, in the context of the M 31 halo, by \citealt{depaolis1995}). The absence of a substantial dependence of the observed temperature asymmetry on the considered \textit{Planck} bands indicates that this asymmetry has to be modulated by the Doppler effect induced by the rotation of the M 90 halo. Option (i) can be excluded since, otherwise, a substantial emission in X-rays should be observable towards the halo of M 90. Option (ii) would require enormous magnetic fields and can be excluded as well. Option (iii) requires a more in-depth analysis but seems unlikely since it would require a  copious supply of dust particles at galactocentric distances up to 300 kpc. It  seems unlikely (even if it cannot be completely excluded at present) that dust grains could be pushed away from the galactic disc by stellar winds and radiation pressure and driven into the M 90 halo. Option (iv) might be active on galaxy cluster scales, where the hot diffuse gas can reach temperatures exceeding a few times $10^7$ K and sizes can be of the order of a few megaparsecs (for details, see e.g. \citealt{cooray2002,chluba2002,altamura2023} and  also \citealt{gurzadyan2005}). In this case, the rkSZ effect due to the ordered hot gas rotation can reach tens of micro-kelvins and constraints have been placed on the rkSZ effect due to the cluster rotation  (see e.g. \citealt{baxter2019}). However, in the case of galactic halos, the effect induced by the rkSZ effect is certainly less important. An in-depth analysis of the M 31 galaxy shows that the rkSZ effect may contribute less than about $0.1 \,\mu$K within a galactocentric distance of about 100 kpc to the temperature asymmetry towards the M 31 halo (see \citealt{tahir2022} and also \citealt {matilla2020}). A similar analysis conducted for M 90 gives a value of $\Delta T/T \simeq 0.2 \,\mu$K within about 100 kpc, assuming a hot gas central number density of about $4\times 10^{-3}$ cm$^{-3}$ and a temperature about $10^6$ K.  Option (v) seems the most likely, as discussed at the end of Sect. 2, and happens in many other spiral galaxies analysed previously using {\it Planck} data.

Another possibility worthy of mention, which presently cannot be excluded, is whether the detected temperature asymmetry can be produced, at least in part, by some hot spot located towards the centre of the Virgo Cluster, that is, towards the M 87 galaxy, which lies about $1.7 \degr$ from M 90 in the A1 quadrant. This possibility certainly deserves further investigation. It seems unlikely that the interaction with M 87 gives rise to a temperature asymmetry with respect to the centre of M 90. On the other hand, it cannot be excluded that M 90 was spun up when it passed near the Virgo Cluster centre, enhancing the temperature asymmetry towards the M 90 halo.  The observed temperature asymmetry would then serve as an empirical test for modelling the survival of a halo upon a galaxy’s passage near the cluster centre.

\begin{acknowledgements}
We acknowledge the use of {\it Planck} data in the Legacy Archive for Microwave Background Data Analysis (LAMBDA) and HEALPix \citep{gorski2005} package. 
We thank for partial support the INFN Projects TAsP (Theoretical Astroparticle Physics) and EUCLID. 
\end{acknowledgements}


\end{document}